\def\itshape{\fontshape\itdefault\selectfont\let\mathrm=\mathit}

\RequirePackage{lineno}

\documentclass[reprint,...]{revtex4-1}
\pdfoutput=1

\usepackage{multirow}
\usepackage{lineno}
\usepackage{graphicx}
\usepackage{epstopdf}
\usepackage{dcolumn} 
\usepackage[retainorgcmds]{IEEEtrantools}
\usepackage{sidecap} 
\usepackage{placeins}
\usepackage{amsmath}
\usepackage{cancel}

\newcommand{\DL}{DAMA/LIBRA }
\newcommand{\SH}{{\it single-hit} }
\newcommand{\MH}{{\it multiple-hit} }
\newcommand{\GS}{LNGS }
\newcommand{\fnu}{\Phi_n^\nu }
\newcommand{\fmu}{\Phi_n^\mu }
\newcommand{\fmumu}{\Phi^{\mu} }
\newcommand{\Rmu}{R_n^\mu }
\newcommand{\Smu}{S^\mu }
\newcommand{\edep}{E_{\text{Dep}} }

\begin{document}

\begin{abstract} 
We present an accurate model of the muon-induced background in the DAMA/LIBRA experiment. Our work challenges
proposed mechanisms which seek to explain the observed DAMA signal modulation with muon-induced backgrounds.
Muon generation and transport are performed using the MUSIC/MUSUN code, and subsequent interactions in the vicinity of the 
DAMA detector cavern are simulated with {\sc Geant4}. 
We estimate the total muon-induced neutron flux in the
detector cavern to be $\Phi_n^\nu = 1.0\times10^{-9}$~cm$^{-2}$~s$^{-1}$.
We predict $3.49\times10^{-5}$~counts/day/kg/keV, which accounts for less than $0.3\%$ of the DAMA signal modulation amplitude. 
\end{abstract}

\title{Muon-induced neutrons do not explain the DAMA data}
\author{J. Klinger, V.~A.~Kudryavtsev}
\address{Department of Physics and Astronomy, University of Sheffield, Sheffield S3 7RH, UK}
\maketitle



\section{Introduction}

The \DL experiment~\cite{Bernabei2008297} is a highly radiopure NaI(Tl) scintillation detector located at the Gran Sasso National Laboratory (LNGS) which
aims to measure the annual modulation signature of dark matter particles~\cite{Bernabei:2009du,doi:10.1142/S0217751X13300226}. Both the \DL experiment 
and the first generation DAMA/NaI experiment
reported the observation of an approximately annual variation in the number of events observed in the 2-6~keV energy range with
a combined significance of approximately 9.3~$\sigma$~\cite{Bernabei:2013xsa}. 
If the observed signal modulation is to be explained by the elastic scattering of dark matter particles, it would require that the interaction cross-section 
of dark matter with nucleons, and the mass of dark matter particles, to be within values that are already excluded by other experiments~\cite{PhysRevD.90.091701,PhysRevLett.106.131302,PhysRevD.86.051701,PhysRevLett.112.091303,PhysRevLett.112.241302,PhysRevLett.107.051301,PhysRevLett.109.181301}.

One mechanism that has been proposed in order to explain the DAMA signal modulation is the production of neutrons due
to the scattering of cosmogenic muons in the material surrounding the detector~\cite{Blum,Ralston,Nygren}. The cosmogenic muon-induced-neutron flux $\fmu$ is expected to have an annual variation related to the mean air temperature above the surface of the Earth that affects the muon flux $\fmumu$ at the surface, and hence underground.
This proposal has been disputed for a number of reasons~\cite{doi:10.1142/S0217751X13300226}, 
notably as the annual variation of cosmogenic muons is approximately 30 days out of phase
with the DAMA signal~\cite{PhysRevD.85.063505,Fernandez-Martinez:1443567,DAMAmuon,DAMAmuon2}.

An extension to this mechanism has also been proposed, which introduces the possibility of a 
contribution to the total neutron flux from the interactions of solar neutrinos~\cite{PhysRevLett.113.081302}. The 
solar neutrino-induced neutron flux~$\fnu$ is also expected to have an annual modulation, due the eccentricity of the Earth's orbit about the Sun. 
It is shown that the phase of $\fnu$ can shift the phase of the total neutron flux relative to $\fmu$.

One should note that most explanations focus on the phase, rather than the amplitude, of the modulation.
A~rough estimate of the modulated rate of muon-induced neutrons~$\Rmu$ in the \DL experiment can be calculated as:
\begin{equation}
\Rmu = \Smu \frac{\fmu A t }{ m} \approx 4.6\times10^{-5} \text{~events~/~day~/~kg}
\end{equation}
where $\fmu$ is taken from previous estimates at \GS\cite{Wulandari:2004bj,Persiani}, $\Smu$ is the amplitude of the muon flux modulation~\cite{selvi2009,Bellini:2012te} 
(relative to $\fmumu$), $A$ is the active surface area perpendicular to $\fmumu$ of one \DL detector module,
 $t$ is the number of seconds in a day and $m$ is the active mass of one detector module.
This rate is three orders of magnitude less than the modulated DAMA signal, even before 
the acceptance for muon-induced events in the DAMA analysis is considered.


Despite the latter calculation, and previous work~\cite{doi:10.1142/S0217751X13300226,DAMAmuon,DAMAmuon2}, the recent paper~\cite{PhysRevLett.113.081302} claims that the 
DAMA signal can partly be explained by muons.
The estimate presented by Davis in Ref.~\cite{PhysRevLett.113.081302}, includes only a very approximate calculation of the amplitude of muon- and neutrino-induced signals. 
The calculation of $\fmu$ contradicts the estimates from Refs.~\cite{doi:10.1142/S0217751X13300226,DAMAmuon,DAMAmuon2}.
The calculation neglects that most neutrons will be accompanied a showering muon, and 
would therefore not be accepted by DAMA/LIBRA. Also, the value of the mean free path for neutrons in rock that is used (taken from Ref.~\cite{PhysRevD.86.054001}), is 
actually for liquid argon; which
has approximately half of the density of \GS rock.
Ref.~\cite{PhysRevLett.113.229001} has additionally shown that the required $\fnu$ is at least six orders of magnitude too large.

It is evident from such contradictions that a full Monte Carlo simulation of the muon-induced background in the \DL experiment is required. 
This would be able to model $\fmu$,
the detector response and the DAMA event acceptance, and also
any enhancement of $\fmu$ due the high-$Z$ shielding used by DAMA/LIBRA.

In this paper, we present an accurate calculation of~$\fmu$ and the total muon-induced background in order to show that the DAMA signal modulation cannot be explained by any muon-induced mechanism.
We perform a full simulation of the \DL apparatus and detector shielding in {\sc Geant4.9.6}~\cite{Agostinelli2003250}, with muons transported to the DAMA cavern 
in \GS using the MUSIC/MUSUN 
code~\cite{Kudryavtsev2009339,Antonioli1997357}. 
In total, we simulate twenty~years of muon-induced data.

\section{Simulation framework}
\label{sim}

We perform the simulation of particle propagation in two stages. In the first stage, only muon transportation from the surface of the Earth down to an underground site 
is considered and secondary particles~\footnote{We define `primary' particles as those which are present at the surface at the Earth, 
and `secondary' particles as those which are subsequently produced.} are neglected. In the second stage, the transport and interactions  of all particles (including secondary particles)
are fully simulated through the material surrounding the \DL apparatus.

\subsection{Muon transport simulation}

The first stage of the simulation is performed using the MUSIC muon transport code~\cite{Kudryavtsev2009339,Antonioli1997357}. The MUSIC code propagates muons from the surface of the Earth 
through a uniform rock of density $\rho = 2.71$~g~cm$^{-3}$ and records energy distributions of muons at different depths. 
The MUSUN code~\cite{Kudryavtsev2009339} calculates muon spectra from the modified Gaisser's 
parameterisation that takes into account the curvature of the Earth and muon lifetime, convoluted with the slant depth distribution at LNGS.
This parameterisation has been previously shown to have a good fit to LVD data~\cite{Kudryavtsev2009339}. 
The MUSUN code subsequently samples muons on the surface of a cuboid with a height of 35~m and perpendicular dimensions  
of 20~m~$\times$~40~m. This cuboid includes most of the corridor where the \DL experiment is located and a few meters of rock around it. 



\subsection{\DL detector simulation}
\label{sim2}

The second stage of the simulation is performed using {\sc Geant4.9.6}~\cite{Agostinelli2003250}.
The {\sc Geant4.9.6} {\sc shielding} physics list has been used, and we include the muon-nuclear interaction process.
The interactions of low-energy neutrons ($< 20$~MeV) are described by high-precision data-driven models~\cite{endf}. 
Previous studies~\cite{Araujo2008471,Persiani,1748-0221-6-05-P05005,Schmidt201328,Reichhart201367} have validated the simulation of neutron production, transport and detection against data.
The level of agreement is better than a factor of two.

In this phase of the simulation,
all primary and secondary particles are transported from the surface of the cuboid until all surviving particles have propagated outside of the cuboid volume. The cuboid is modeled as \GS rock with a density $\rho = 2.71$~g~cm$^{-3}$ and a chemical composition as described in Ref.~\cite{Wulandari2004313}.
A corridor (`cavern') is positioned within the cuboid,
such that there is 10~m of \GS rock overburden, and 
otherwise 5~m of \GS rock surrounding the cavern walls and floor. 

The \DL detector housing is placed halfway along the length of the cavern, adjacent to a cavern wall. The housing is composed of \GS concrete
with density  $\rho = 2.50$~g~cm$^{-3}$ and a chemical composition as described in Ref.~\cite{Wulandari2004313}. 
The \DL apparatus and detector housing are described in Ref.~\cite{Bernabei2008297}.
There are a number of concentric layers of shielding surrounding the \DL detector. Extending outwards from the detector, we model
10~cm of copper, 15~cm of lead, 1.5~mm of cadmium, 50~cm of polyethylene and 1~m of \GS concrete.

We model each of the 25 \DL detector modules, 
containing a central cuboidal crystal composed of NaI, in addition to light-guides and photomultiplier tubes~\cite{Bernabei2008297}.
The dimensions of each module, including a further 2~mm of copper shielding,
are $10.6\times10.6\times66.2$~cm$^3$. The 25 modules are placed in a $5\times5$ arrangement 
in the vertical and width dimensions of the cavern.
%

\section{The muon-induced neutron flux}
\label{flux}

\begin{table}[b]
\squeezetable
\renewcommand{\arraystretch}{1.2} 
\caption{A comparison of $\fmu$ (in units of $10^{-10}$~cm$^{-2}$~s$^{-1}$) predicted by this study, Wulandari et al.~\cite{Wulandari:2004bj} and Persiani~\cite{Persiani}.
The column titled `Cavern' indicates the three distinct cavern geometries used: (1)~in this study; (2) by Wulandari et al. and (3) by Persiani. The range of considered 
neutron energies is shown, and `(*)' indicates that back-scattered neutrons are included.}
\centering
\begin{tabular}{lcccc}
\toprule 
& Cavern & $> 0$~MeV & $> 1$~MeV & $> 1$~MeV (\bf{*})  \\
\colrule 
This study          & (1) & 10 & 4.0 & 5.0 \\
This study          & (2) & 7.6 & 5.8 & 10 \\
Wulandari et al. & (2) & No data & 4.3 & 8.5 \\
Persiani             & (3) & 7.2 & 2.7 & No data \\
\botrule
\end{tabular}
\label{tab:nuflux}
\end{table}

In the stage of simulation described in Section~\ref{sim2}, muons and secondary particles are transported through 10~m of \GS rock above the DAMA cavern, 
and also through 5~m on the sides and underside of the cavern. Integrating over the surface area of the cavern, our simulation predicts $\fmu = 1.0\times10^{-9}$~cm$^{-2}$~s$^{-1}$,
excluding 
back-scattering~\footnote{We define the back-scattered neutron flux as including independent counts from neutrons that re-enter the cavern due to scattering in the surrounding rock.}. 

In Table~\ref{tab:nuflux}, we compare our result to the simulation of Wulandari et al.~\cite{Wulandari:2004bj}, 
which is performed 
using FLUKA~\cite{fluka}, and of Persiani~\cite{Persiani}, which is performed using {\sc Geant4.9.3} and MUSIC/MUSUN. 
Integrating over all neutron energies, our results are consistent within about 30\% of the previous estimates.

We additionally demonstrate a dependance of $\fmu$ on the dimensions of the cavern
by scaling $\fmu$ to the cavern proportions used by Wulandari et al. (compare the first two rows in Table~\ref{tab:nuflux}).
We attribute this to the different fluxes and energy spectra of vertical and inclined muons.

\begin{figure}[t]
  \centering
  \includegraphics[width=\linewidth]{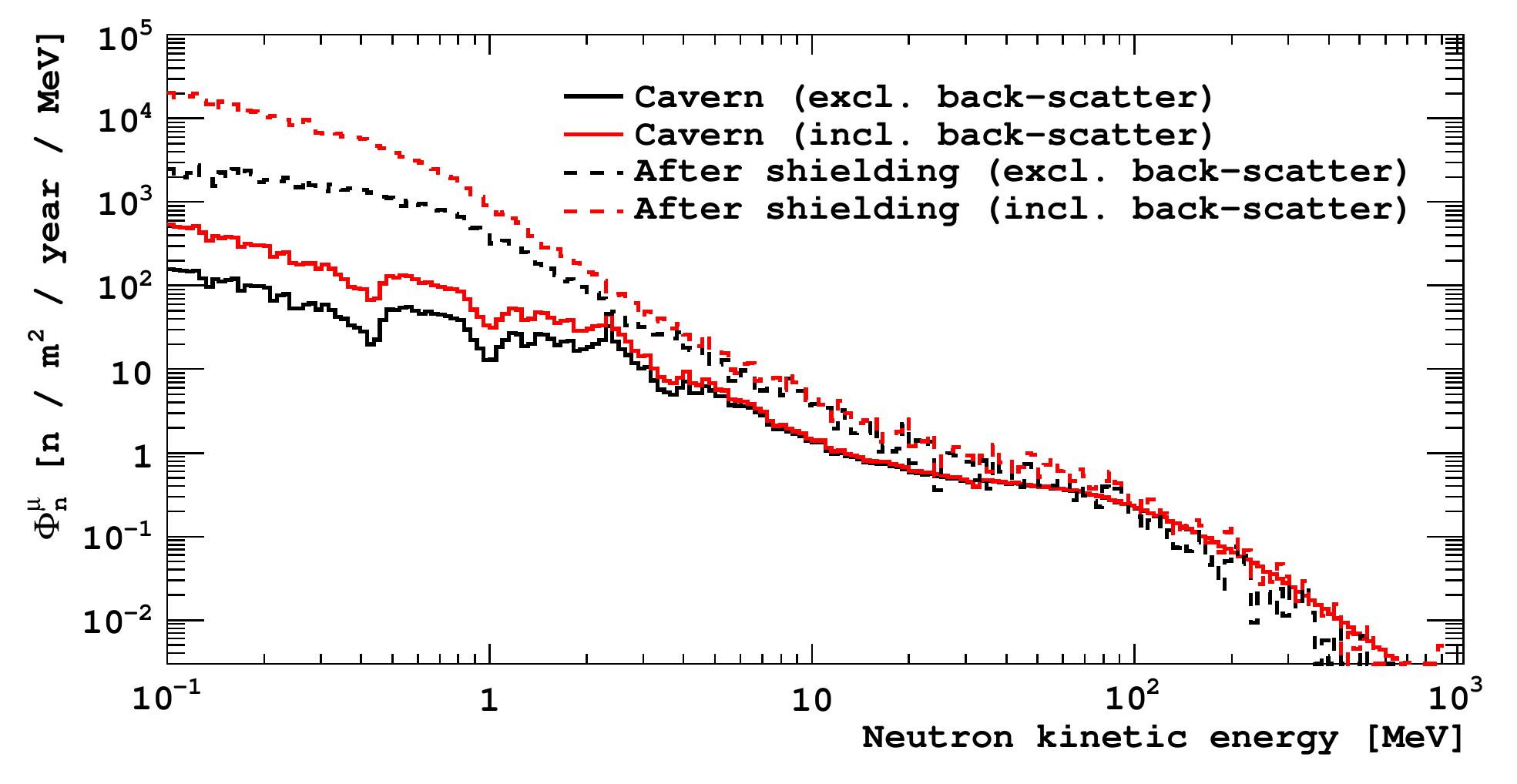}
  \caption{The distribution of $\fmu$ as a function of neutron energy for neutrons entering the cavern in which \DL is situated (solid lines) and
  entering the \DL detector modules after all shielding is traversed (dashed line). The distributions excluding and including back-scattered neutrons are shown in black and red
  respectively.}
  \label{fig:neutronFlux}
\end{figure}

High-$Z$ materials in the detector shielding (lead and copper) will lead to an enhancement of $\fmu$ which could, potentially,
 contribute to the modulated signal.
Figure~\ref{fig:neutronFlux} shows $\fmu$ as a function of neutron energy, as predicted in the \GS cavern and after 
all particles are propagated through the various layers of the \DL shielding. It is shown that $\fmu$ increases by a factor of $> 5$ due to the shielding.
As we will discuss in Section~\ref{results}, this enhancement of $\fmu$ is still insufficient to explain the DAMA data.

\section{Analysis}
\label{results}

In our simulation, each detector module is treated independently and all information due to an energy deposition from any particle in the NaI crystal volumes is recorded.
DAMA categorises events as being \SH or \MH if the event has an energy deposit in only a single crystal or in multiple crystals
respectively. The DAMA signal region, in which the modulated signal is observed, is then defined only for \SH events with a total energy deposit ($\edep$) in the range 2-6~keV.

In the following sections, we will show that 
the number of muon-induced events entering the \DL signal region is too low to explain the signal modulation.

\subsection{Resolution and quenching factors}

\begin{figure}[t]
  \centering
\includegraphics[width=\linewidth]{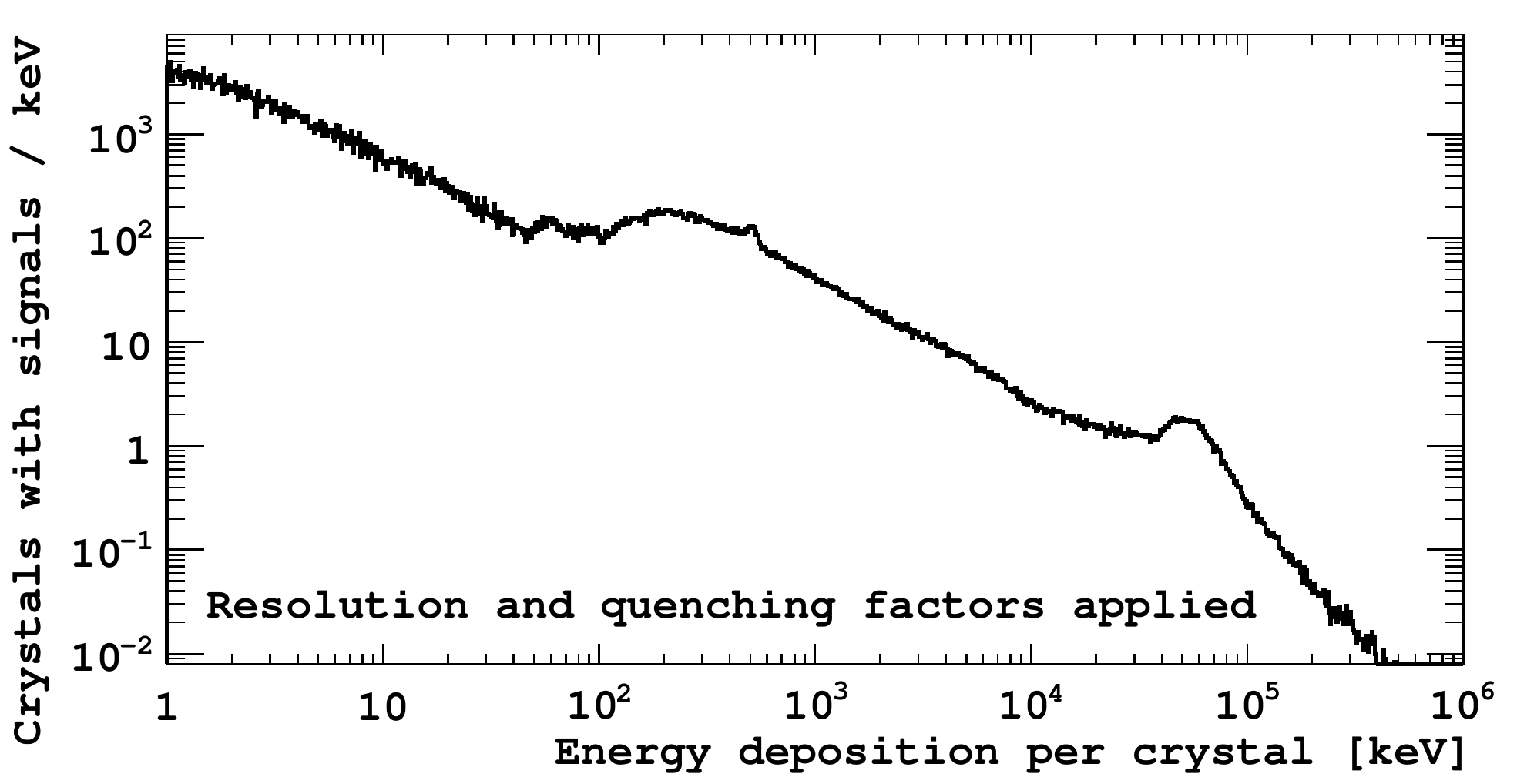}
  \caption{The energy spectrum for all crystals with energy depositions, with resolution and nuclear recoil quenching factors applied.
  The equivalent of twenty years of muon-induced data is presented.}
  \label{fig:neutronenergy}
\end{figure}

We model the \DL experimental resolution by applying a Gaussian smearing to the sum of all energy deposited in each crystal, using resolution parameters
provided by Ref.~\cite{Bernabei2008297}. 
As {\sc Geant4} does not account for the quenching of energy depositions in nuclear recoils,
we apply correction factors obtained from previous
studies~\cite{Tovey:1998ex,Spooner:1994ca,1748-0221-3-06-P06003,Gerbier:1998dm,Simon:2002cw}. 
Figure~\ref{fig:neutronenergy} shows the energy spectrum of all crystals with energy
depositions, after corrections factors are applied.

\subsection{Single-hit and multiple-hit events}
\label{singlehit}

\begin{figure}[t]
  \centering
\includegraphics[width=\linewidth]{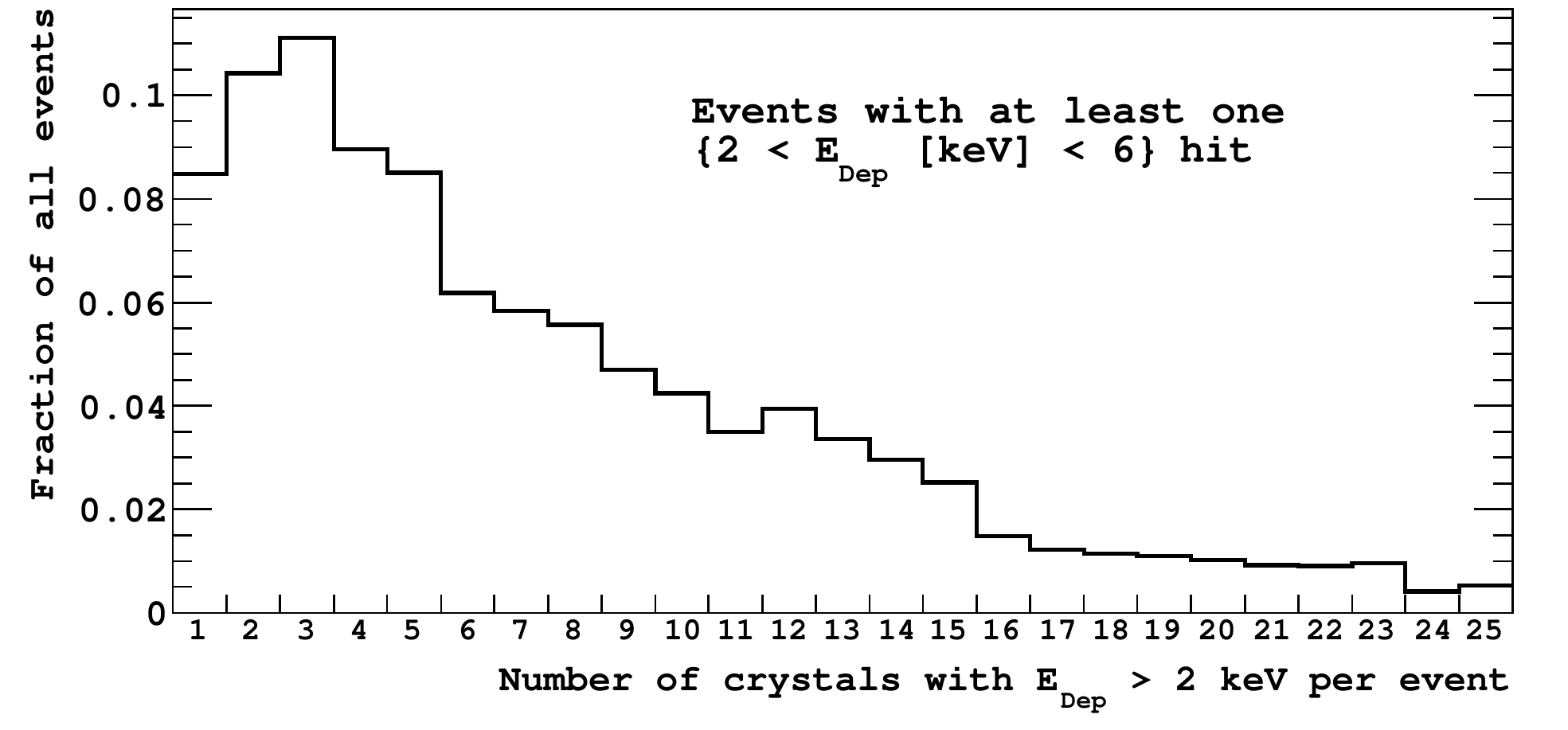}
  \caption{The distribution of the hit multiplicity in events with $\edep \geq 2$~keV per crystal.
  At least one crystal has a total energy absorption in the range 2-6~keV. The equivalent of twenty years of muon-induced data is presented.}
  \label{fig:ncrystals}
\end{figure}

An important detail that is neglected in previous attempts~\cite{Blum,Ralston,Nygren,PhysRevLett.113.081302} to explain the DAMA signal modulation with muon-induced 
backgrounds is the 
acceptance for \SH events. Figure~\ref{fig:ncrystals} shows the number of crystals in events with 
$\edep \geq 2$~keV per crystal, and in which
at least one crystal in the event has a total energy absorption of 2-6~keV. It is shown that~$< 9\%$ of these events are \SH events, which suppresses
 any enhancement of $\fmu$ due to interactions in the detector shielding.

\FloatBarrier

\subsection{Events in the signal region}
\label{signalregion}

\begin{figure}[t]
  \centering
\includegraphics[width=\linewidth]{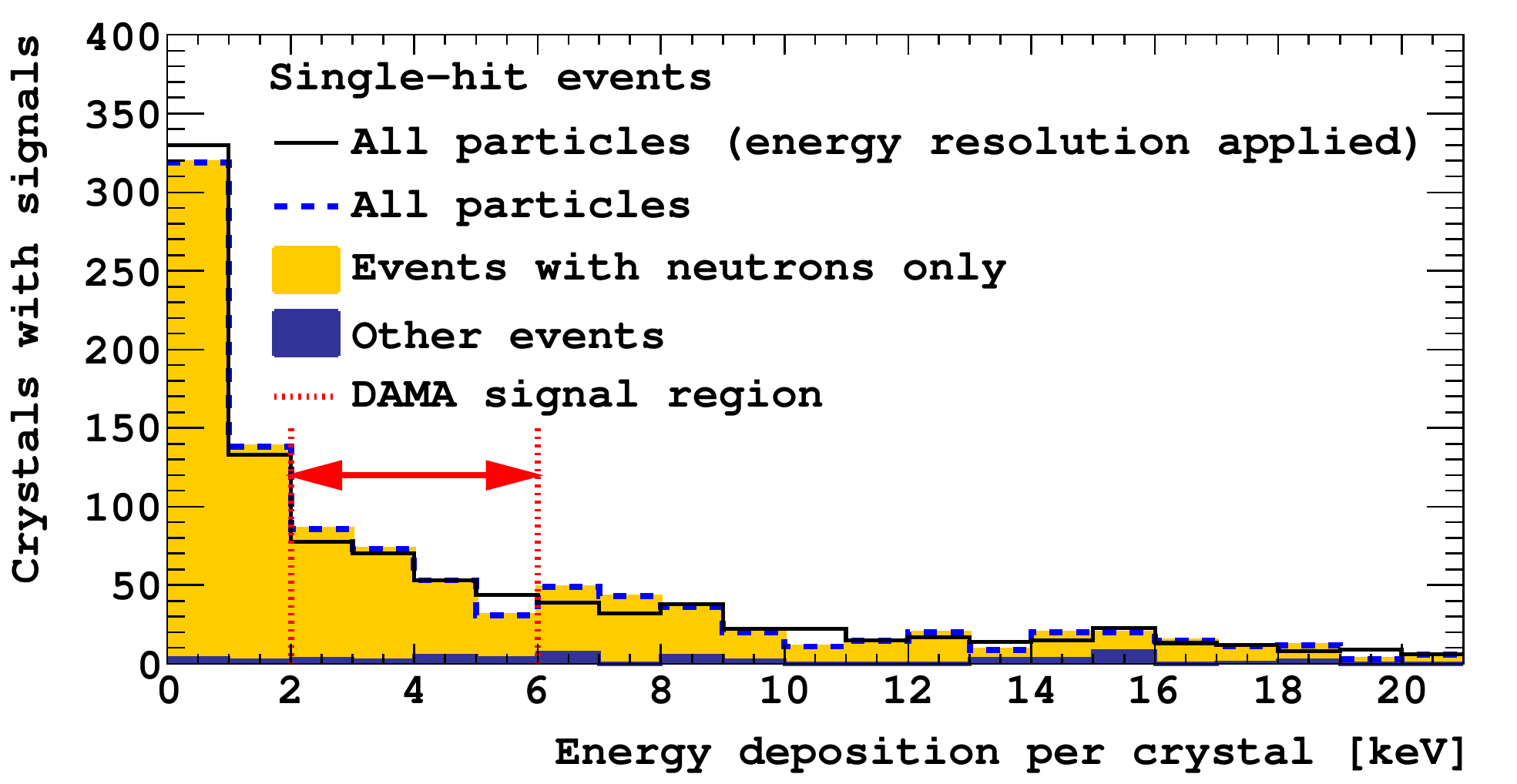}
  \caption{The distribution of the total energy deposition in \SH events. The blue dashed line and the black solid line show the sum
  of all energy depositions
  before and after energy resolution is considered, respectively. The stacked colored bars indicate the relative fraction of 
  all events (before energy resolution is considered) attributed to events in which only neutrons deposit energy (yellow) and other events (blue). The equivalent of twenty years of muon-induced data is presented.}
  \label{sig:signalregion}
\end{figure}

In this section, we present the number of muon-induced \SH events predicted by our simulation. The distribution
of the energy deposited in crystals in \SH events is shown in Figure~\ref{sig:signalregion}. For $\edep < 20$~keV, the muon-induced background 
is dominated by isolated neutrons. 

In the range 2-6~keV there are 245 muon-induced events predicted over a period equivalent to twenty years. 
The total sensitive mass of the \DL detector is 242.5~kg, therefore we predict the rate of muon-induced events in this energy range
 to be $3.49\times10^{-5}$ counts~/~day~/~kg~/~keV with approximately 6\% statistical uncertainty. We estimate the systematic uncertainty to be
  approximately~30\% by comparing different predictions of $\fmu$, as presented in Section~\ref{flux}.
 
 We are able to compare our prediction to the conservative estimate presented in Ref.~\cite{DAMAmuon,DAMAmuon2} which is in agreement with our results.
 
 The calculated event rate accounts for $\sim 0.3\%$ of the modulation amplitude reported by DAMA of 
 $(1.12 \pm 0.12)\times10^{-2}$~counts~/~day~/~kg~/~keV~\cite{Bernabei:2013xsa}.
 It is clear from this comparison that, even if the systematic uncertainty is bigger than our estimates, no muon-induced background can be used to explain the observed signal modulation. 
 Our simulations (Figures~\ref{fig:ncrystals} and~\ref{sig:signalregion}) also show that, if muon-induced backgrounds could explain the DAMA data, 
 one should expect a non-negligible modulation of the muon-induced background above 6~keV, as well as for events with multiple hits, which is not seen by DAMA.

\section{Discussion}

In this section we will argue that muon-induced neutrons cannot explain the DAMA data, 
even before any estimate of $\fmu$ is performed.

We start the discussion in a general way, 
by considering any possible source of modulated signal, including dark matter, as has been done in Ref.~\cite{Kudryavtsev201091}. 
The measured rate of events at \DL is clearly dominated by radioactive background above 6 keV, which imposes a strict limit on any interpretation of the modulated signal. 
This radioactive background is almost flat at low energies~\cite{Kudryavtsev201091}, with the 
exception of a peak from $^{40}$K at about 3~keV, which agrees with the DAMA measurements. 
To preserve the shape (`flatness') of the radioactive background in the region 2-6~keV, 
the total signal should be small and hence, the modulated fraction of the signal should be large.
 As an example, the measured modulated signal rate of 0.019~counts~/~day~/~kg~/~keV at 2-3~keV, assumed to be 5\% of the total (average) signal, 
 will give the total signal rate of 0.38~counts~/~day~/~kg~/~keV. This 
 is already a significant fraction of the total measured rate at 2-3~keV (about 30\%),
 requiring the radioactive background rate to drop by 30\% at this energy whilst maintaining a flat background above 6~keV.
 No model of radioactivity 
predicts a dip in the background below 6~keV~\cite{Kudryavtsev201091}.

Let us now consider muon-induced backgrounds within this context.
We assume that $\fmu$ and $\fmumu$ are modulated in a similar way,
linked to the mean muon energy at \GS\cite{Kudryavtsev2003688}.
The LVD~\cite{selvi2009} and Borexino~\cite{Bellini:2012te} experiments have observed a muon flux modulation in the range of 1.3-1.5\% of the total $\fmumu$.
If the modulated signal in DAMA is due to a muon-induced effect, then the total rate of 
this `effect' will be $0.0112 / 0.014 \approx 0.8$~counts~/~day~/~kg~/~keV. This is 
approximately equal to the total rate of $\sim 1$~counts~/~day~/~kg~/~keV observed by DAMA in the 2-6~keV energy range~\cite{Bernabei:2013xsa}. 
The effect is more dramatic in the 2-3 keV energy range, where the modulated signal is approximately 0.0190~counts~/~day~/~kg~/~keV~\cite{Bernabei:2013xsa}.
This would imply a total muon-induced background of~$0.0190 / 0.014 \approx 1.4$~counts~/~day~/~kg~/~keV, which is higher that the total rate of 
events observed by DAMA. 
This is excluded by radioactivity models~\cite{Kudryavtsev201091}. 

It is clear from the latter discussion that for any explanation of the DAMA signal to be consistent with the measured spectrum of events,
it must satisfy the following qualitative criteria:

\begin{itemize}
\item The amplitude of the effect must be very small compared to the DAMA event rate.
\item The modulation amplitude of the effect must not be much smaller than the average amplitude of the effect.
\item Any effect not satisfying the latter two criteria implies that there is a new model of suppressed radioactivity in the region 2-6 keV, that does not apply above 6 keV.
\item The modulation of the effect must only affect \SH events, whilst disregarding \MH events.
\item The explanation must simultaneously predict the phase and the period of the modulation.
\end{itemize}

\noindent An explanation which incorporates muon-induced backgrounds cannot satisfy these criteria.

\section{Conclusions}

We have presented an accurate simulation of the muon-induced background in the \DL experiment, in response to 
proposals to explain the observed DAMA signal modulation with muon-induced neutrons.
We have performed a full simulation of the \DL apparatus, shielding and detector housing using {\sc Geant4.9.6}. 

We have calculated the muon-induced neutron flux in \GS to be $\fmu = 1.0\times10^{-9}$~cm$^{-2}$~s$^{-1}$ (without back-scattering), which is consistent
with previous simulations. After selecting events which satisfy the DAMA signal region criteria, our simulation predicts a background rate
of $3.49\times10^{-5}$~counts~/~day~/~kg~/~keV. 
This accounts for approximately $0.3\%$ of the modulation amplitude. 
We find that one would expect a non-negligible modulation of muon-induced background above 6~keV, as well as for events with multiple hits, which is not seen by DAMA.

We conclude from our study that muon-induced neutrons are unable explain the DAMA data. 
Furthermore, a large signal event rate, independently of the source of this signal, is inconsistent with radioactive background models.

\FloatBarrier

\bibliography{muneu-dama-v5}

\end{document}